\documentclass[a4paper]{article}

\usepackage{INTERSPEECH2021}

\usepackage[utf8]{inputenc}

\usepackage{url}

\title{Extending Text-to-Speech Synthesis with Articulatory Movement Prediction using Ultrasound Tongue Imaging}
\name{Tamás Gábor Csapó$^{1,2}$}
\address{
  $^1$Department of Telecommunications and Media Informatics, \\
	Budapest University of Technology and Economics, Budapest, Hungary \\
	$^2$MTA-ELTE Lendület Lingual Articulation Research Group, Budapest, Hungary}
\email{csapot@tmit.bme.hu}

\begin{document}

\maketitle
\begin{abstract}
In this paper, we present our first experiments in text-to-articulation prediction, using ultrasound tongue image targets. We extend a traditional (vocoder-based) DNN-TTS framework with predicting PCA-compressed ultrasound images, of which the continuous tongue motion can be reconstructed in synchrony with synthesized speech. We use the data of eight speakers, train fully connected and recurrent neural networks, and show that FC-DNNs are more suitable for the prediction of sequential data than LSTMs, in case of limited training data. Objective experiments and visualized predictions show that the proposed solution is feasible and the generated ultrasound videos are close to natural tongue movement. Articulatory movement prediction from text input can be useful for audiovisual speech synthesis or computer-assisted pronunciation training.   
\end{abstract}

\section{Introduction}

Statistical parametric methods are frequently used in text-to-speech (TTS) synthesis, with two main machine learning techniques: hidden Markov-models (HMM, \cite{Zen2007}) and deep neural networks (DNN, \cite{Zen2013}). Recently, the focus of TTS research has moved to end-to-end solutions, applying  neural vocoders (e.g.~WaveNet~\cite{Oord2016} and WaveGlow~\cite{Prenger2019}) and sequence-to-sequence models using attention (e.g.~Tacotron2~\cite{Shen2018}). Still, traditional (non-end-to-end, vocoder-based) DNN-TTS systems are useful in limited data scenarios, when there is few training data available, for example with speech and biosignal recordings, or in case of audiovisual speech synthesis.

\subsection{Audiovisual speech synthesis}

Audiovisual speech synthesis is a subfield of the more general areas of speech synthesis and computer facial animation \cite{Massaro2015}. The goal of the visible speech synthesis is typically to obtain a mask with realistic motions, not to duplicate the musculature of the face to control this mask. 

The field of visual speech synthesis is fairly well established and a variety of approaches have been developed (including rule-based \cite{Perrier2014} and data-driven methods \cite{Schabus2013}). Rule-based systems include models for speech sequence planning, for muscle mechanisms and for the physical speech production apparatus. Within the biomechanical model of the vocal tract, the tongue can be represented as a finite element mesh \cite{Perrier2014} and complex biomechanical simulations are necessary to estimate the internal muscle stresses during the movement of human articulators \cite{Stavness2014}.
In the context of data-driven approaches and HMM-based synthesis, there are two main categories: image-based systems are supposed to look like a video of a real person, while motion capture based systems derive features from facial points tracked over time \cite{Schabus2013}. 
For HMM-based audiovisual synthesis, a synchronous corpus of parametrized facial motion data and acoustic speech data is necessary. 
Schabus et al. showed that in combined HMM-based text-to-speech synthesis and facial animation, joint audiovisual models perform better than training separate acoustic and visual models \cite{Schabus2013}.


\subsection{Predicting articulatory movement from text}

Another type of TTS extension is when the target is to predict articulatory motion (e.g.~lip or tongue movement) and not just the face of the speaker, besides the speech output. This requires special biosignals to be recorded, which can track the movement of the articulatory organs (e.g.\ EMA, X-ray, vocal tract MRI, and ultrasound tongue imaging). With such a system, by giving an arbitrary input text, one is able to hear the speech and, in synchrony with it, see how to move the tongue in 3D to produce target speech sounds. This visual feedback can make a big difference for pronunciation training in L2 learning, especially when the target language contains speech sounds which are difficult to articulate.

Most earlier studies in this context were using point-tracking devices, like electromagnetic articulography (EMA)~\cite{Ling2010,Ling2010a,Wei2016a,Steiner2017,LeMaguer2017,Yu2018}. Ling and his colleagues proposed a HMM-based text-to-articulatory movement prediction system, i.e.\ which can synthesize the speaker's mouth from text~\cite{Ling2010}. Here, the durations were not modeled, but in a subsequent study, they also investigated the timing aspects and analyzed the critical articulators~\cite{Ling2010a}. Wei et al.\ used DNNs for the text-to-EMA prediction and confirmed that stacked bottleneck features are effective for this purpose~\cite{Wei2016a}. Steiner and his colleagues experimented similarly with text-to-EMA prediction using HMMs (with synchronous text-to-speech), and  they also included a geometric 3D tongue model as the target~\cite{Steiner2017}. Next, they compared HMMs and DNNs for the text-to-tongue model prediction~\cite{LeMaguer2017}. It was found that with less than 2 hours of data, DNNs outperformed HMMs. Yu and her colleagues predicted 3D articulatory movement, from text and audio inputs, therefore combining the text-to-speech and acoustic-to-articulatory inversion fields~\cite{Yu2018}. For the machine learning approach, they used a bottleneck long-term recurrent convolutional neural network. They showed that the text information complements well the acoustic features during the prediction of EMA-based articulation. The final output of the system is speech synchronized with 3D articulatory animation, using a facial mesh model~\cite{Yu2018}.







As shown above, there have been several studies investigating text-to-articulatory motion with HMMs or DNNs, but all of them are using point-tracking equipment (electromagnetic articulography). Medical imaging target, like ultrasound or MRI, have not been used before in this context.

\begin{figure*}
\centering
\includegraphics[trim=0.0cm 25.4cm 0.0cm 0.0cm, clip=true, width=0.9\textwidth]{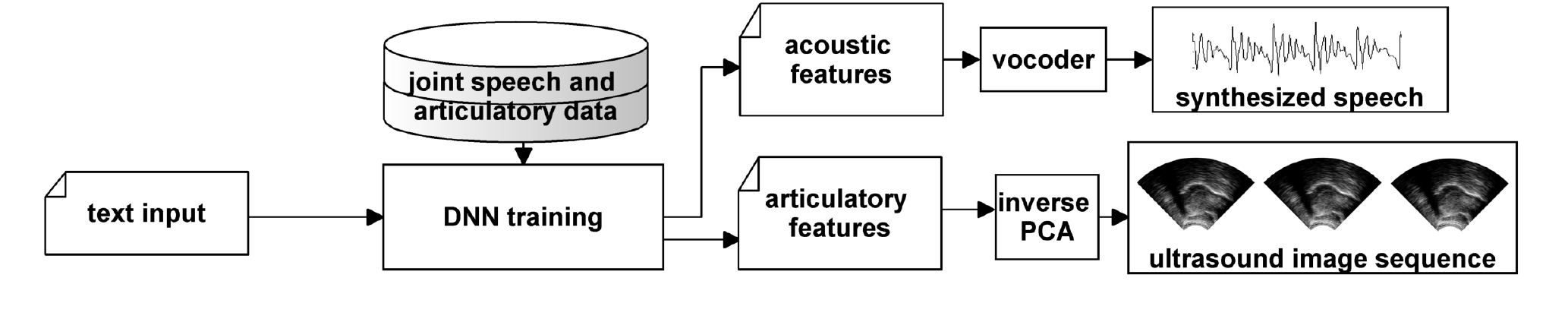}

\caption{Block diagram of the proposed approach.}
\label{fig:txt2ult}
\end{figure*}

\subsection{Ultrasound tongue imaging}

Ultrasound tongue imaging (UTI) is a technique suitable for the acquisition of articulatory data. Phonetic research has employed 2D ultrasound for a number of years for investigating tongue movements during speech \cite{Stone1983}. Stone summarized the typical methodology of investigating speech production using ultrasound~\cite{Stone2005a}. Usually, when the subject is speaking, the ultrasound transducer is placed below the chin, resulting in mid-sagittal images of the tongue movement. The typical result of 2D ultrasound recordings is a series of gray-scale images in which the tongue surface contour has a greater brightness than the surrounding tissue and air.
Compared to other articulatory acquisition methods (e.g.\ EMA, X-ray, XRMB, and vocal tract MRI), UTI has the advantage that the tongue surface is fully visible, and ultrasound can be recorded in a non-invasive way~\cite{Stone2005a,Csapo2017c,Ramanarayanan2018}. An ultrasound device is easy to handle and move, since it is small and light, and thus it is suitable for fieldworks, as well. Besides, it is a significantly less expensive piece of equipment than the above mentioned devices.

In our earlier studies, we have shown that ultrasound is a feasible solution for articulatory-to-acoustic mapping~\cite{Csapo2017c,Csapo2020c} and acoustic-to-articulatory inversion~\cite{Porras2019}. However, text-to-ultrasound synthesis has not been investigated before.

\subsection{Contributions of this paper}

The goal of this paper is to extend DNN-TTS with articulatory movement prediction, using ultrasound images of the tongue. We show on the data of several speakers that the combined TTS and synthesized articulatory motion is feasible and can result in acceptable articulatory movement video. Text-to-articulatory movement prediction might be useful for computer-assisted pronunciation training (CAPT) applications and articulatory visualization.

\section{Methods}

\subsection{Data}

For the data, we used the UltraSuite-TaL80 database~\cite{Ribeiro2021} (\url{https://ultrasuite.github.io/data/tal_corpus/}). We chose four English male (03mn, 04me, 05ms, 07me) and four female speakers (01fi, 02fe, 06fe, and 09fe). In parallel with speech, the tongue movement was recorded in midsagittal orientation using the ``Micro'' ultrasound system of Articulate Instruments Ltd.\ at 81.5~fps. Lip video was also recorded in UltraSuite-TaL80, but we did not use that information in the current study. The ultrasound data and the audio signals were synchronized using the tools provided by Articulate Instruments Ltd. Each speaker read roughly 200 sentences -- the duration of the recordings was about 15 minutes, which was partitioned into training, validation and test sets in a 85-10-5 ratio.

\subsection{Processing the ultrasound data}

In our experiments, articulatory features estimated from the raw scanline data of the ultrasound were used as the additional target of the networks (see Fig.~\ref{fig:txt2ult}). The  64$\times$842 pixel images were resized to 64$\times$128 pixels using bicubic interpolation, and we calculated PCA coefficients, similarly to EigenTongues~\cite{Hueber2007}. While calculating the PCA, we aimed at keeping the 70\% of the variance of the original images, thus having 128 coefficients. An example for the PCA eigenvectors can be seen in Fig.~\ref{fig:ultpca}, and the result of PCA is presented in Fig.~\ref{fig:ultpca_pred_video}. To be in synchrony with the acoustic features (frame shift of 5~ms), the ultrasound data was resampled to 200~Hz.

\subsection{DNN-TTS framework}

Fig.~\ref{fig:txt2ult} illustrates the proposed approach, i.e.~the combined acoustic and articulatory feature prediction using a DNN from text input.
The experiments were conducted in the Merlin DNN-TTS framework~\cite{Wu2016} (\url{https://github.com/CSTR-Edinburgh/merlin}). Textual / phonetic parameters are first converted to a sequence of linguistic features as input (based on a decision tree).
Next, neural networks are employed to predict acoustic (60-dimensional MGC, 5-dimensional BAP, and 1-dimensional LF0, with delta and delta-delta features) and articulatory features (128-dimensional ULT-PCA, with delta and delta-delta) as output for synthesizing speech, at a 5~ms frame step with the WORLD vocoder. From the predicted 128-dimensional articulatory features, the 64$\times$128 image is reconstructed using the PCA matrix, and bicubic interpolation is applied to resize the image to 64$\times$842 pixels, to be comparable with the original data. For visualization purposes, we transformed this raw scanline data to wedge format, which shows how the real aspect ratio of the tongue surface (for an example, see Fig.~\ref{fig:ultpca_pred_video}). The transformation was done with 'ultrasuite-tools' (\url{https://github.com/UltraSuite/ultrasuite-tools}) 

\subsubsection{FC-DNN}

The DNN used here is a fully-connected feed-forward multilayer perceptron architecture (FC-DNN, six hidden layers, 1024 neurons in each). We applied tangent hyperbolic activation function, SGD optimizer, and a batch size of 256. The input features had min-max normalization, while output acoustic features had mean-variance normalization. We trained the networks for 25 epochs with a warm-up of 10 epochs, applying early stopping, and a learning rate of 0.002 after that with exponential decay. We only trained both a duration model and an acoustic model, the latter also containing the articulatory features.

\subsubsection{LSTM}

Recurrent networks are typically more suitable to process sequential data. Therefore, we also trained an LSTM network following the Merlin recipe (four FF layers with 1024 neurons each, and one LSTM layer with 512 neurons). To ensure a longer training with the recurrent network, we used ADAM optimizer, and a warm-up of 30 epochs with early stopping. The other parameters were the same as for the FF-DDN. We trained both duration and acoustic models.

All neural network trainings were done individually with each speakers' data, without average voice training or adaptation.


\section{Experimental results}

After training the above models, we synthesized sentences from the test parts of the ultrasound datasets. To measure the validation and test error, we calculated both spectral prediction error (Mel-Cepstral Distortion, MCD), and an articulatory feature related error (ULT-PCA/RMSE, calculated on the normalized PCA values). We trained both duration and acoustic models, but for the error calculations, we synthesized the test sentences with their original timing. This way, warping the features in time was not necessary for calculating the error measures.

\subsection{Demonstration samples}

\begin{figure*}
\centering
\includegraphics[trim=0.0cm 0.0cm 0.0cm 0.0cm, clip=true, width=0.9\textwidth]{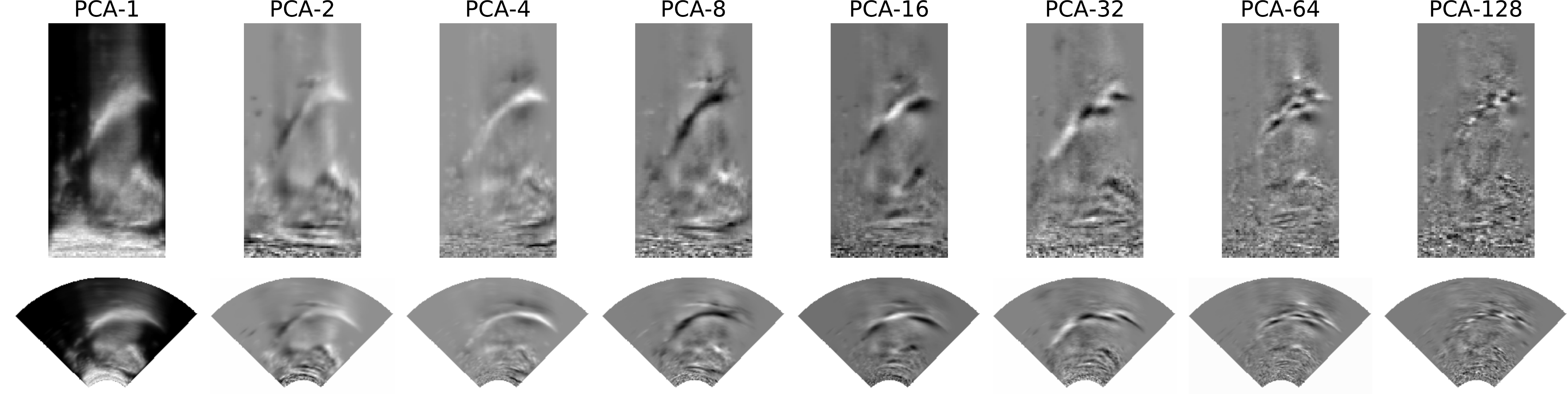}

\caption{PCA eigenvectors, from speaker '01fi'. Top: raw, scanline data (resized to 64$\times$128 pixels). Bottom: wedge orientation.}
\label{fig:ultpca}
\end{figure*}

\begin{figure*}
\centering
\includegraphics[trim=0.0cm 0.5cm 0.0cm 1.2cm, clip=true, width=0.9\textwidth]{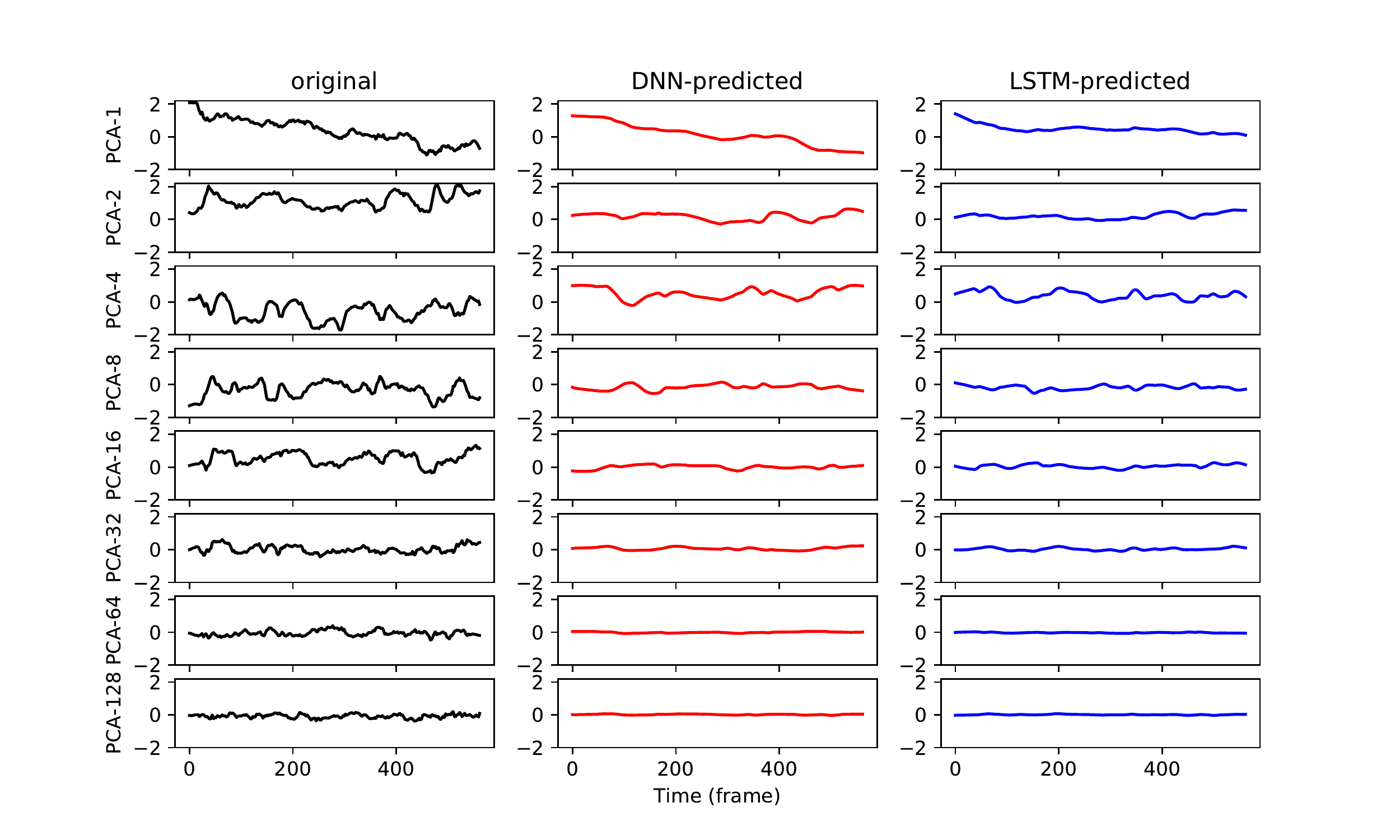}

\caption{Original and predicted articulatory features, from speaker '01fi'. Sentence: '"I leave it to nobody," said Shakespeare, sulkily.'}
\label{fig:ultpca_pred}
\end{figure*}

An example for the PCA eigenvectors are in Fig.~\ref{fig:ultpca}, and the predicted articulatory feature sequence can be seen in Fig.~\ref{fig:ultpca_pred} (speaker '01fi', sentence '201\_aud'). As lower dimensional PCA vectors contain more information, the visualization was done in an exponential way and only 8 dimensions are plotted out of 128. Clearly, both the FC-DNN and the LSTM are following the tendencies found in the original data (e.g.\ in case of PCA-1, PCA-2, PCA-4), but the fine details are not modeled well. This type of oversmoothing is a frequent phenomena in statistical parametric speech synthesis. The higher dimensions (e.g.\ PCA-64 and PCA-128) are almost constant; i.e.\ they could not be modeled well with neural networks.

\begin{figure*}
\centering
\includegraphics[trim=0.0cm 0.0cm 0.0cm 0.0cm, clip=true, width=\textwidth]{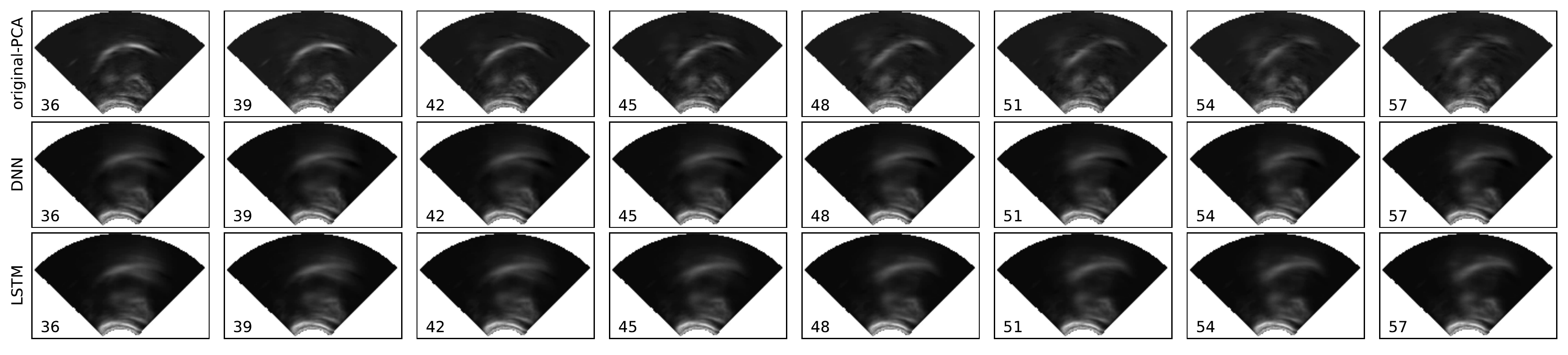}

\vspace{4mm}

\includegraphics[trim=0.0cm 0.0cm 0.0cm 0.0cm, clip=true, width=\textwidth]{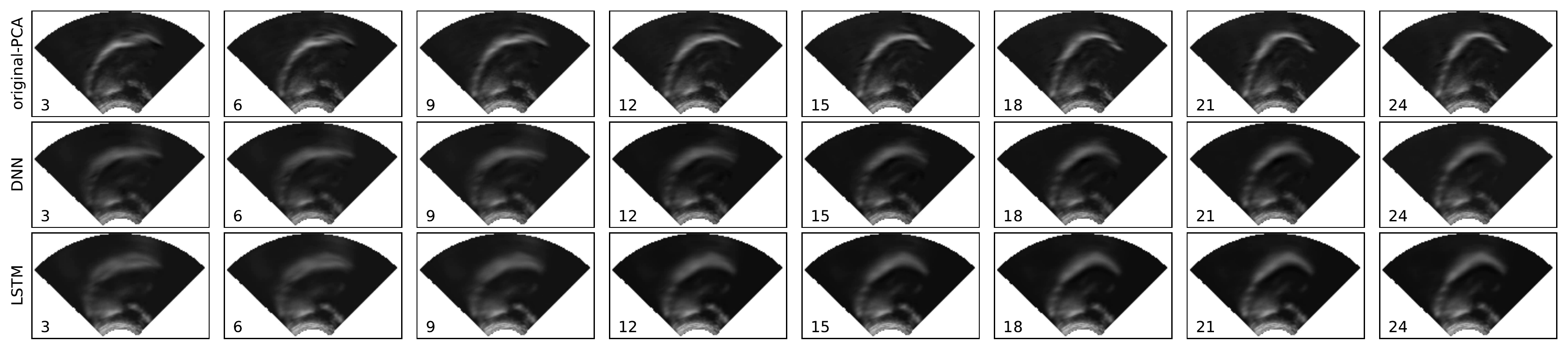}

\caption{Original and predicted articulatory feature sequences. Top: speaker '01fi', bottom: speaker '03mn'. The numbers at the bottom-left correspond to the frame number of the video.}
\label{fig:ultpca_pred_video}
\end{figure*}

To visualize the individual ultrasound images, we plotted several ultrasound frames from the original videos and from the predicted ones, as a function of time, in Fig.~\ref{fig:ultpca_pred_video}. The reason why we are plotting every third frame is that for the 5~ms frame step of the Merlin toolkit, the 81.5~fps ultrasound video was interpolated to 200~Hz, and therefore, in the predicted data, roughly every 3rd frame contains visible tongue motion. In case of speaker '01fi', we can see in the top row (original ultrasound images after PCA) that there is a significant tongue movement, i.e.~the tongue tip (on the right) goes higher, as the time passes. Both the predictions with the DNN and LSTM network follow the articulatory movement, but the images are smoothed -- again, resulting from the statistical training. For speaker '03mn', similar tendencies can be observed: the movement of the tongue is changing its curvature as a function of time, but in the DNN-predicted and LSTM-predicted images, the tongue surface is not as clear as in the original data.

As the synthesized motion of the tongue is more visible in real-time videos, we made available several samples at \newline \url{http://smartlab.tmit.bme.hu/ssw11_txt2ult}.

\subsection{Objective measures}

Table~\ref{tab:objective_MCD} summarizes the MCD results. The MCD values of the test sentences with the FC-DNN are between 5.8--7.0~dB (average: 6.228~dB), whereas with LSTM they are between 6.0--7.5~dB (average: 6.593~dB), indicating that the recurrent neural network was not helpful in estimating the acoustic features. The reason for this might be that we have limited articulatory-acoustic databases (roughly 200 sentences for each speaker), which is too small for training an LSTM model.

The results of the RMSE calculated on the articulatory feature are summarized in Table~\ref{tab:objective_ULTPCA128}. The lowest error was achieved with the data of speaker 09fe: with FC-DNN, the test error is 2.9, while with LSTM, the test error is 3.1. The tendency is similar to the case of MCD: the LSTM network was not helpful in predicting the articulatory features, probably due to the small size of the data.

\begin{table}
\caption{MCD errors on the dev/test set.} \label{tab:objective_MCD}
\centering
\begin{tabular}{l||c|c}
     & \multicolumn{2}{c}{{MCD}} \\
Speaker & FC-DNN & LSTM \\
\hline\hline
01fi & 6.995  / 6.971  & 6.647  / 6.588  \\
02fe & 6.095  / 5.803  & 6.486  / 6.259  \\
03mn & 5.781  / 5.785  & 5.977  / 5.948  \\
04me & 5.896  / 6.024  & 6.318  / 6.312  \\
05ms & 6.244  / 6.256  & 7.235  / 7.083  \\
06fe & 5.758  / 5.582  & 6.444  / 6.330  \\
07me & 6.589  / 6.562  & 6.831  / 6.749  \\
09fe & 6.516  / 6.844  & 7.197  / 7.472  \\
\hline
average  & 6.234  / 6.228  & 6.642  / 6.593  \\
\hline
\end{tabular}
\end{table}

\begin{table}
\caption{ULTPCA/RMSE errors on the dev/test set.} \label{tab:objective_ULTPCA128}
\centering
\begin{tabular}{l||c|c}
     & \multicolumn{2}{c}{{ULTPCA128/RMSE}} \\
Speaker & FC-DNN & LSTM \\
\hline\hline
01fi & 3.292  / 3.223  & 3.319  / 3.208  \\
02fe & 3.533  / 3.732  & 3.753  / 3.904  \\
03mn & 3.147  / 3.660  & 3.289  / 3.680  \\
04me & 3.849  / 3.985  & 4.031  / 4.033  \\
05ms & 3.133  / 3.233  & 3.249  / 3.405  \\
06fe & 3.439  / 3.250  & 3.743  / 3.451  \\
07me & 3.544  / 3.595  & 3.498  / 3.461  \\
09fe & 3.022  / 2.864  & 3.234  / 3.133  \\
\hline
average  & 3.370  / 3.443 & 3.515  / 3.534  \\
\hline
\end{tabular}
\end{table}


\section{Discussion and conclusions}

We have shown above that text-to-ultrasound video prediction is feasible as an extension to traditional DNN-based text-to-speech synthesis, despite the relatively small amount of training data. Although the synchrony between visual and speech output is not enforced by the model, the tied acoustic and articulatory features during the DNN training ensure that the audio and visual features are in synchrony, i.e.~that in the generated ultrasound videos, the tongue is moving according to the synthesized speech. To objectively check this, SyncNet, part of Wav2Lip could be applied to assess synchrony~\cite{Prajwal2020}. Our paper found that the joint learning of both acoustic and articulatory features has advantages, but this is not substantiated -- a comparison of the joint model against two separate models remains future work. 

Although there have been several earlier attempts for extending text-to-speech synthesis with articulatory data, all of these studies were using EMA, being a point tracking equipment~\cite{Ling2010,Ling2010a,Wei2016a,Steiner2017,LeMaguer2017,Yu2018}, and containing less spatial information about the tongue than ultrasound. The advantage of ultrasound in this context is that the resulting video shows a larger portion of the tongue, compared to EMA.

Articulatory movement prediction from text input can be useful for audiovisual speech synthesis. A specific application is computer-assisted pronunciation training / computer-aided language learning~\cite{Katz2014,Jones2017,Agarwal2019}, which can be beneficial for learners of second languages. With such a combined TTS and text-to-articulatory prediction system, by giving an arbitrary input text, one is able to hear the speech and, in synchrony with it, see how to move the tongue in 2D or 3D to produce target speech sounds. This visual feedback can be helpful for pronunciation training in L2 learning, especially when the target language contains speech sounds which are difficult to articulate.

In the future, we plan to investigate speaker adaptation and speaker-independent training. For this, a common articulatory space has to be defined, as the currently used PCA representation is specific for each individual speaker. Also, multi-task learning might be useful in this context: a system could potentially be pre-trained on speech-only material which is easier to acquire, and the UTI be trained only in addition. Besides, we plan to investigate the effect of the misalignments in the ultrasound transducer position~\cite{Csapo2020d,Csapo2020e} on the text-to-ultrasound prediction results. 

The code is accessible at \url{https://github.com/BME-SmartLab/txt2ult}.

\section{Acknowledgements}

The author was partly funded by the National Research, Development and Innovation Office of Hungary (FK 124584 and PD 127915 grants). We would like to thank CSTR for providing the Merlin toolkit and the UltraSuite-TaL articulatory database.


\bibliographystyle{IEEEtran}

\bibliography{ref_collection_csapot_nourl}

\end{document}